# A look into the future of the COVID-19 pandemic in Europe: an expert consultation

Emil Nafis Iftekhar, Max Planck Institute for Dynamics and Self-Organization, Göttingen, Germany
Viola Priesemann, Max Planck Institute for Dynamics and Self-Organization, Göttingen, Germany
Rudi Balling, University of Luxembourg, Luxembourg, Luxembourg
Simon Bauer, Max Planck Institute for Dynamics and Self-Organization, Göttingen, Germany
Philippe Beutels, University of Antwerp, Antwerp, Belgium
André Calero Valdez, RWTH Aachen University, Aachen, Germany
Sarah Cuschieri, University of Malta, Msida, Malta
Thomas Czypionka, Institute for Advanced Studies, Vienna, Austria, and London School of Economics, London, UK
Uga Dumpis, Pauls Stradins Clinical University Hospital, University of Latvia, Riga, Latvia
Enrico Glaab, University of Luxembourg, Luxembourg, Luxembourg
Eva Grill, Ludwig-Maximilians-University München, München, Germany
Claudia Hanson, Karolinska Institute, Stockholm, Sweden, and London School of Hygiene & Tropical Medicine, London, UK
Pirta Hotulainen, Minerva Foundation Institute for Medical Research, Helsinki, Finland
Peter Klimek, Medical University of Vienna, Vienna, Austria and Complexity Science Hub Vienna, Vienna, Austria
Mirjam Kretzschmar, University Medical Center Utrecht, Utrecht, The Netherlands
Tyll Krüger, Wroclaw University of Science and Technology, Wroclaw, Poland
Jenny Krutzinna, University of Bergen, Bergen, Norway
Nicola Low, University of Bern, Bern, Switzerland
Helena Machado, Institute for Social Sciences, University of Minho, Braga, Portugal
Carlos Martins, Department of Community Medicine, Health Information and Decision Sciences of the Faculty of Medicine, University of Porto, Porto, Portugal
Martin McKee, London School of Hygiene & Tropical Medicine, London, UK
Sebastian Bernd Mohr, Max Planck Institute for Dynamics and Self-Organization, Göttingen, Germany
Armin Nassehi, Ludwig-Maximilians-University München, München, Germany
Matjaž Perc, University of Maribor, Maribor, Slovenia, and Department of Medical Research, China Medical University Hospital, China Medical University, Taichung, Taiwan
Elena Petelos, University of Crete, Crete, Greece, and Maastricht University, Maastricht, The Netherlands
Martyn Pickersgill, University of Edinburgh, Edinburgh, UK
Barbara Prainsack, Department of Political Science, University of Vienna, Vienna, Austria
Joacim Rocklöv, Department of Public Health and Clinical Medicine, Section of Sustainable Health, Umeå University, Umeå, Sweden
Eva Schernhammer, Medical University of Vienna, Vienna, Austria
Anthony Staines, Dublin City University, Dublin, Ireland
Ewa Szczurek, University of Warsaw, Warsaw, Poland
Sotirios Tsiodras, National and Kapodistrian University of Athens, Athens, Greece
Steven Van Gucht, Sciensano, Brussels, Belgium
Peter Willeit, Medical University of Innsbruck, Innsbruck, Austria, and University of Cambridge, Cambridge, UK

# Abstract

How will the coronavirus disease 2019 (COVID-19) pandemic develop in the coming months and years? Based on an expert survey, we examine key aspects that are likely to influence COVID-19 in Europe. The future challenges and developments will strongly depend on the progress of national and global vaccination programs, the emergence and spread of variants of concern (VOCs), and public responses to nonpharmaceutical interventions (NPIs). In the short term, many people are still unvaccinated, VOCs continue to emerge and spread, and mobility and population mixing is expected to increase over the summer. Therefore, policies that lift restrictions too much and too early risk another damaging wave. This challenge remains despite the reduced opportunities for transmission due to vaccination progress and reduced indoor mixing in the summer. In autumn 2021, increased indoor activity might accelerate the spread again, but a necessary reintroduction of NPIs might be too slow. The incidence may strongly rise again, possibly filling intensive care units, if vaccination levels are not high enough. A moderate, adaptive level of NPIs will thus remain necessary. These epidemiological aspects are put into perspective with the economic, social, and health-related consequences and thereby provide a holistic perspective on the future of COVID-19.

# Main text

## Introduction

More than a year after the World Health Organization declared the coronavirus disease 2019 (COVID-19) a Public Health Emergency of International Concern, Europe continues to struggle with it. Although future developments are highly uncertain, we aim to provide (a) a systematic assessment of the factors that will affect the course of the COVID-19 pandemic in Europe, and (b) a tentative forecast of how the pandemic may evolve prior to coming to an end in Europe. We chose a method inspired by the Delphi method of forecasting[1] as the most suitable way to elicit expert opinions about key developments and themes regarding the COVID-19 pandemic. The facilitators developed questionnaires with open-ended questions and asked scientists from various European countries, disciplines, and research fields, to provide their input and predictions. As the guiding questionnaires were focussed on epidemiology, virology, public health, and social science, some other important perspectives, such as those of clinical medicine, economics, and the humanities, are not covered in great detail (see SI). Here we set out the results of the expert consultation - outlining salient commonalities and divergent responses. Of necessity, this paper represents a partial synthesis of the rich and diverse contributions, and not all authors necessarily agree in detail with every single statement.

We first summarize insights on three critical factors that shape the development of the epidemic: population immunity and vaccination, variants of concern (VOCs), and public responses to pandemic policy. Second, we present scenarios based on the available knowledge as of April 2021 for three distinct time periods: for (a) summer 2021, (b) autumn and winter 2021, and (c) for a period of 3-5 years from spring 2021. For the latter period, we give a high-level overview of the consequences of the COVID-19 pandemic for health, society, and the economy. In the last section, we elaborate in more detail on central topics mentioned in the main text: long-term strategy, vaccination coverage, organization of mass vaccinations, waning immunity, evolution of the virus, improving adherence, airborne transmission, and One

Health. We hope that the insights of our synthesis will serve as a scientific basis for policy debates by generating a comprehensive overview of key considerations in moving beyond the pandemic, while informing other foresight studies.

## Key factors determining the course of the pandemic

Our starting point is the situation as of spring 2021. During the COVID-19 waves in winter 2020-2021, many European countries experienced high numbers of infections that, in some places, overwhelmed hospitals. This was partly due to insufficient ICU capacity in some countries.[2] Delayed responses and lower effectiveness of non-pharmaceutical interventions (NPIs) compared to the first wave also played a part.[3] Even countries that have had relatively few cases and a low death toll until then were hit severely in the winter. As of early 2021, Europe is experiencing another surge in cases, which appears to have peaked in April 2021. The emergence and severity of these waves has varied greatly across Europe (see Figure 1 and 2). The future development of the pandemic will also likely be heterogeneous. In the following sections we focus on three key factors that contribute to this heterogeneity.

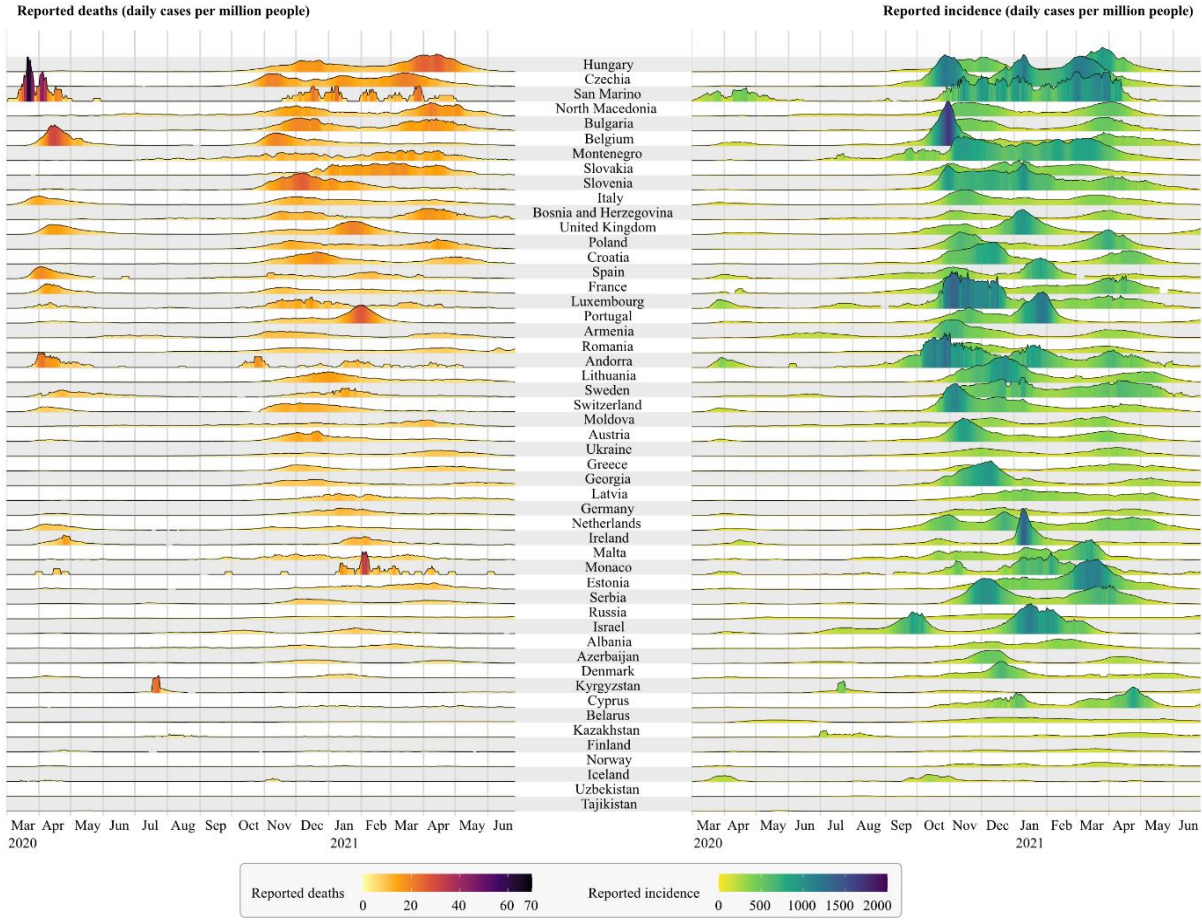

Figure 1: Comparison of the COVID-19 pandemic in all countries of the WHO European Region (except for Turkey and Turkmenistan as there was no appropriate data available in the data set). Countries are ordered from top to bottom with a decreasing cumulative number of COVID-19 related deaths per million people. The y-axis scale of the ridgeline plots is the same for all countries for reported deaths and incidence, respectively. Even though reported numbers are associated with wide uncertainty, the differences between countries and waves are evident. Data source: https://corona-api.com (Accessed: June 28, 2021).

## Population immunity and vaccination

Population immunity (also referred to as herd immunity) describes a situation in which enough people in the population are immune to a pathogen, such that it is not able to spread widely (WHO, 2020a). The proportion of immune people in the population needed to reach population immunity in a given country is mainly driven by the infectivity of severe acute respiratory syndrome coronavirus 2 (SARS-CoV-2) and the ability of either past natural infection or vaccines to reduce transmission.[4] Models that assume basic reproduction numbers of 2·5-3·5 have previously estimated that transmission-blocking immunity of 60-72% of the population is required in the case of SARS-CoV-2.[5, 6] This figure is higher for more transmissible variants. Therefore a *minimum* immunization level of 80% of the entire population is likely to be required.[7, 8] This figure would be difficult to achieve with vaccination alone if vaccines are not fully protective against infection or prevent onward transmission. Furthermore, immunization needs to be homogeneous across all population groups, otherwise pockets of transmission can prevail. To achieve this goal, one might consider mandatory vaccinations - the effectiveness of which remains contested, as vaccination uptake depends on a complex interplay of different factors.[9, 10] A 2016 systematic review found that mandatory childhood vaccination policies were associated with improved uptake[11], a finding supported by later experience in Italy.[12, 13] However, there are many legal, ethical, cultural, and technical issues involved and it has been argued that it should only be considered when all other reasons for low uptake, such as accessibility, have been addressed and the decision should take account of the particular context and the risk of unintended consequences.[9, 14-17] In any case, for the short term it is more important to distribute available vaccines to locations where they are most needed.[18]

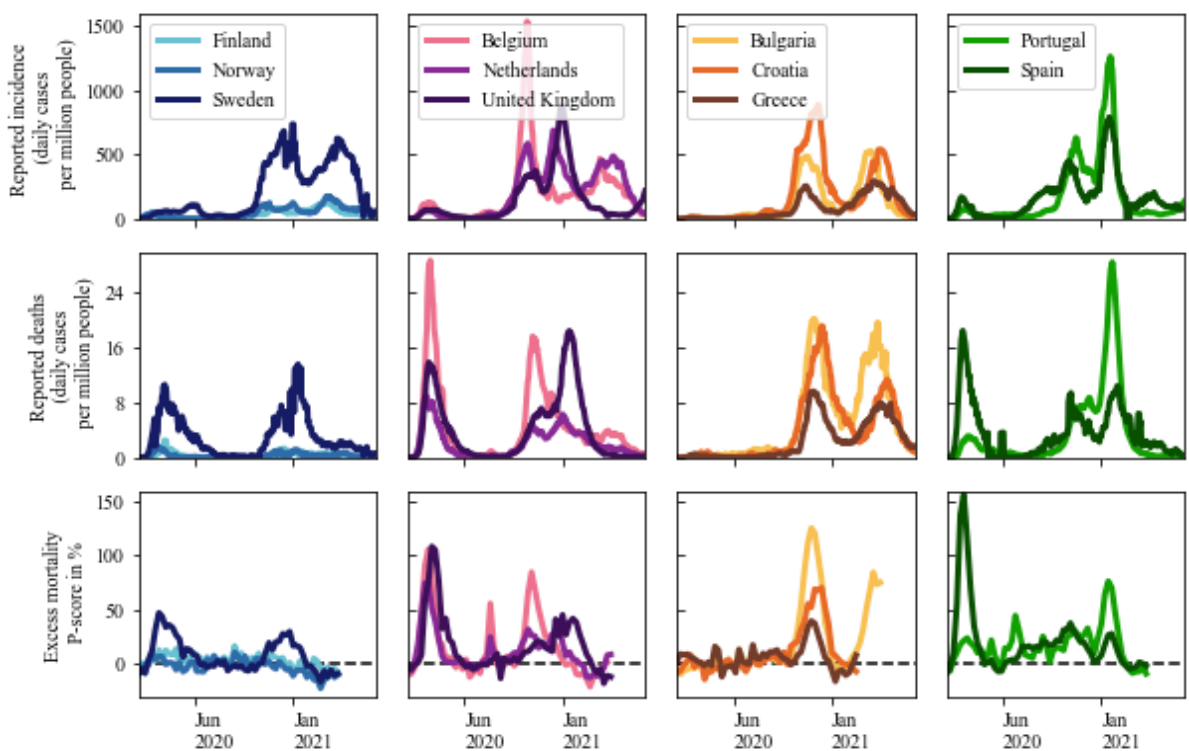

Figure 2: Comparison of the COVID-19 pandemic in a selection of European countries grouped by geographical proximity. Many differences in reported incidence, reported deaths and excess mortality can be observed. Even though reported numbers are associated with wide uncertainty, the differences

between countries and waves are evident. Data sources: https://ourworldindata.org/covid-cases and https://ourworldindata.org/excess-mortality-covid (Accessed: June 29, 2021).

One contribution to population immunity comes from so-called natural immunity, as a result of prior infection with SARS-CoV-2 and potentially by cross-immunity due to prior exposure to other coronaviruses.[19, 20] The fraction of those who are naturally immune in the population varies widely between European countries. However, in all countries the majority of the population remained susceptible to infection.[21]

In individuals who have had a SARS-CoV-2 infection, antibodies have been shown to persist for up to nine months after infection.[22] About 95% of people retain immune memory at six months after infection.[23-25] This indicates that the likelihood of reinfection and severe disease progression is low in this time frame, but reinfection is still possible.[26-28]

The second, major, contributor to population immunity is vaccination. The first vaccines are, as of April 2021, licensed for use in adults and the vaccines appear to reduce infections by varying amounts, typically in the 80-90% range for mRNA vaccines (after two doses)[29-31] and potentially lower for others.[32, 33] Vaccines are, however, still likely to reduce transmissibility even if breakthrough infection occurs.[34] Importantly, they seem especially likely to prevent severe symptoms and hospitalization, reaching relative risk reductions of about 70-95%.[30, 32, 35-37] The progress of vaccination programs is continuing in Europe (see Figure 3).[38]

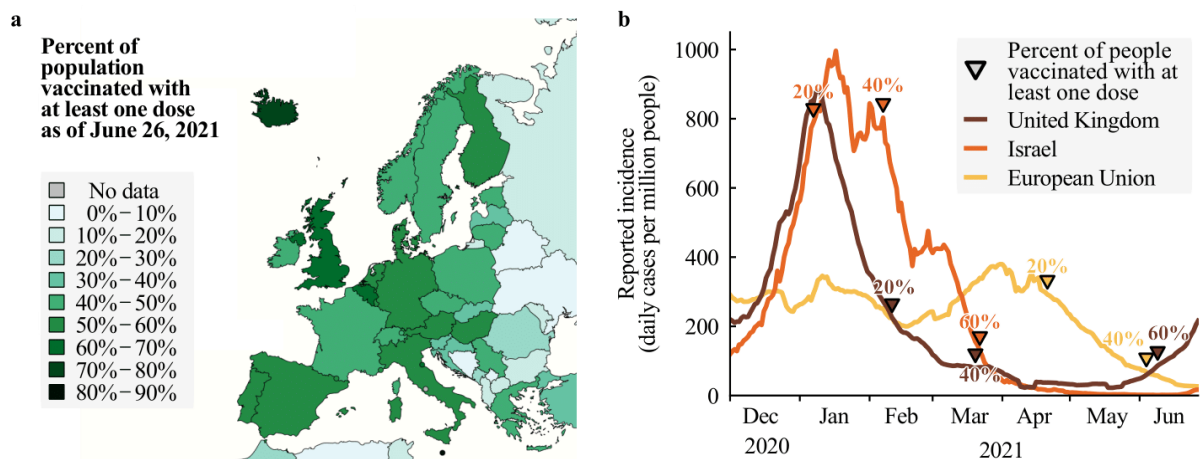

Figure 3: Vaccination progress in Europe. **a.** Fraction of the population having received at least one dose of COVID-19 vaccines in Europe as of June 26, 2021. There are large differences in vaccination coverage. **b.** Reported incidence (lines) and reached vaccination milestones (triangles) since the start of vaccination programs. Data source: https://ourworldindata.org/covid-vaccinations (Accessed: June 29, 2021).

The chances of achieving high vaccination coverage depend on a multitude of factors including political leadership, trust in public health and other public authorities, access to and eligibility for vaccines, and vaccine acceptance. The last is especially crucial. As of April 2021, acceptance is lower for the non-mRNA-vaccines with lower reported efficacies. Repeatedly changing policy recommendations and constant media coverage further unsettled people, especially after evidence of possible links to rare adverse, sometimes fatal, side-effects emerged mid-rollout for the AZD1222 (AstraZeneca) and Ad26.COV2.S (Johnson & Johnson) vaccines.[39, 40] Among older people and the most vulnerable, who have been receiving the vaccine in the initial phase, vaccine uptake has been generally high.[41, 42] In younger age groups, willingness to get vaccinated appears lower[43, 44] – in France, only about 40% of the working age population currently plan to accept a vaccine.[45] Moreover, vaccine uptake in the groups of healthcare workers is rather disconcerting in some countries - e.g., Belgium and France - has been low.[46-48] However, perception of increasing vaccine uptake might motivate those who are hesitant.[49] To conclude, the issue of vaccine uptake presents an ever-changing situation.[50]

## Variants of concern

VOCs are so called because they harbour certain mutations that have consequences for SARS-CoV-2 pathogenicity. Existing and newly emerging SARS-CoV-2 VOCs are challenging because, compared to the original variant, they may increase transmissibility or severity, prolong the duration of the infectious period, shorten the duration of post-infection immunity, or escape host immune responses to natural infection or to vaccines. They could also affect diagnostic testing accuracy, the spectrum of detectable symptoms, and therapeutic management. The frequency and the spectrum of variants of SARS-CoV-2 will depend on functional constraints and evolutionary pressure.

The Alpha (B.1.1.7) variant, which was first detected in the United Kingdom, demonstrated enhanced transmissibility[51, 52], a longer duration of acute infection[53], a higher hospitalization rate[54], and probably a higher infection fatality rate than previously circulating variants.[51, 55-57] The Beta (B.1.351) variant, which was first detected in South Africa, exhibits higher transmissibility[58], while the impact on disease severity of this variant remains uncertain as of April 2021.[59] The Beta and Gamma (P.1) variants, the latter originated in Brazil, seem to partially evade the immune response of previously infected individuals.[26, 60] In Europe, the Alpha variant became the dominant variant in December/January 2020 in, e.g., the UK, Ireland and Portugal, and in February/March 2021 in, e.g., France and Germany.[61] In contrast, the Beta and Gamma variants have not become widely distributed in Europe so far. TheDelta (B.1.617.2) variant appears to be more transmissible than previous strains.[62]

There is uncertainty about the efficacy of available vaccines in relation to VOCs. Current vaccines appear to be effective against Alpha.[29, 31] However, there is some evidence that the efficacy of some vaccines might be reduced for Beta, Gamma, and Delta.[32, 62-64] It remains unclear to which degree this is the case, and how much the protection against severe courses of disease might be affected.

The more infections are present in the human population, the higher the rate of mutation. This can lead to selection for VOCs with transmission advantage or, in places with high rates of natural or vaccinal immunity, VOCs with escape mutations. In countries without well established genetic surveillance, this may permit uncontrolled spread. In this case, vaccines will need to be updated to protect against these new VOCs, with the consequent requirements to gain approval, be manufactured, and distributed anew. However, the more widespread infection is, the more mutations will occur that could end up with an evolutionary advantage. Consequently, the best safeguard is to reduce transmission. Only after sufficient global vaccination coverage will the mutation rate decrease due to lower viral spread in the post-pandemic phase.[8]

## Public responses to pandemic policy

As long as population immunity has not been reached, maintaining appropriate and widely accepted levels of NPIs to mitigate the spread remains crucial.[65, 66] When there is a rise in infections, NPIs must be reimplemented or strengthened; the earlier this is done, the more effective it is.[67] However, the resoluteness and timeliness with which NPIs are being implemented and remain in place depends on leadership and public opinion.[68] Moreover, the higher the efficacy of NPIs the more the public accept and support them.[69]

As of spring 2021, pandemic policies are not being received well in many parts of Europe.[70] A range of factors likely contribute to this, including continued high economic[71-73] and psychological burdens[74-79], inadequate risk communication[80-83], the lack of transparent long-term strategies from governments[68], increasing vaccination coverage (see Figure 3) and a general erosion of trust.[84-88] All this results in lower adherence to rules and recommendations for mitigating the spread of SARS-CoV-2 compared to the first wave.[70, 89]

The effectiveness of rules and recommendations depends on the ability and willingness of the population to adhere to them.[81] Adherence in the past year has varied from country to country. In some countries, adherence was initially quite high in general.[89-93] In others, there have been strong protests against measures, sometimes resulting in their relaxation.[94-96] In general, voluntary adherence will be more likely if the necessity for and strategy behind instituted measures is communicated clearly and systematically, and if interpersonal trust and public trust in government is higher.[70, 97-100] However, if COVID-19-induced morbidity and mortality reaches levels that societies deem intolerable, acceptance of NPIs rises again[70]

Given these key factors underlying the future evolution of the pandemic, we can consider what to expect in the future, beginning with the summer of 2021.

## The perspective for the summer of 2021

Summer 2021 is likely to bring some relief in Europe as people spend more time outside[101], vaccination proceeds, and control strategies improve, e.g., via improved availability and variety of testing technology.[102] The expected relief might be compromised if the combination of natural immunity and vaccination coverage is low and relaxation of NPIs is not managed carefully. Furthermore, increased international travel will increase the risk of importing any VOCs that emerge from outside of Europe, and the risk of circulating any VOCs that emerge from within the continent across European nations. If VOCs with an ability to evade immune responses emerge, NPIs may need to be reinstated or strengthened even in populations where relatively high levels of immunity have been achieved. A common European goal to keep infection levels low and to internationally coordinate close surveillance of incidence and viral genomes, especially of infected international travelers, would help to reduce the risk of emergence of VOCs.[103]

Once vaccination coverage is deemed sufficiently high by decision makers, countries might come under further pressure to ease measures again. With (most) risk groups vaccinated first, there will be a lower fraction of severe illnesses and deaths related to COVID-19 in the population. Consequently, a lower burden on healthcare systems is also expected. However, some individuals at risk might not (yet) have been vaccinated, protection by vaccination is not perfect and may wane over time, and unvaccinated and possibly some vaccinated people will continue to transmit. This makes it unlikely that restrictions can be lifted *completely* without risking another larger wave. Another wave would result in increased morbidity and mortality of unvaccinated people, or in general those to whom the vaccines did not confer protection.[104] With vaccine strategies first targeting older people, a wave in summer would predominantly hit relatively younger age groups. It would also further strain exhausted healthcare personnel and healthcare systems now functioning beyond capacity for protracted periods of time. Hence, certain mitigation strategies will need to remain in place in an adaptive manner.[105] When considering retaining NPIs, countries might also take the opportunity to achieve low case numbers as, with increasing immunization, the containment of COVID-19 is facilitated. In a situation of low case numbers, an effective test-trace-and-isolate (TTI) system, supported by digital contact tracing apps, further facilitates epidemic control.[106] In such a regime, only a few NPIs, such as wearing (FFP2) masks or basic hygiene measures, might have to stay in place.

To summarise, in the summer of 2021, countries could still be faced with overwhelmed intensive care units and ongoing strict imposition of NPIs. This is a consequence of the limits of the vaccines available, inadequate vaccination coverage, increased mobility across borders and regions, and the possibility of escape variants. However, if a country succeeds in maintaining low case numbers and slows down the influx and spread of any new VOC with sound epidemiological surveillance and reactive measures, then moderately strict NPIs similar to those in summer 2020, or potentially even fewer restrictions, may be possible. The exact extent of NPIs that are necessary to prevent an overburdening of health systems regionally depends on various factors, such as the characteristics of prevalent VOCs and vaccination coverage. A full lifting of all restrictions (e.g., for large indoor gatherings), however, is unlikely to be possible in summer 2021 without risking further outbreaks.

## The perspective for the autumn and winter of 2021

What can be expected in the autumn and winter of 2021 depends substantially on what happens in the summer; specifically, the success of vaccination programs both in Europe and worldwide, and the emergence and spread of (new) VOCs. Compared to the summer, autumn and winter bring the additional complication of unfavorable seasonal effects.

The seasonality of coronaviruses is expected to increase infections in the autumn and winter months[101, 107, 108], with increased indoor contacts.[109] Additionally, other seasonal viruses, such as influenza and respiratory syncytial virus, could cause more pressure on health services than in 2020. Since there might be fewer restrictions, and possibly lower-than-usual levels of population immunity because one season of transmission was "skipped", these other seasonal viruses are likely to circulate in greater numbers than in 2020 .[110, 111] Overall, the transition to autumn and winter could be problematic because restrictions might have to be tightened again to prevent a rapid rise in case numbers. Based on experiences in several European states in autumn and winter 2020-2021, there is a risk that reintroduction of the necessary public health measures may come too late to succeed in preventing another wave in autumn. It will be the task of governments not to repeat these mistakes.

In the *best-case* scenario, vaccination efforts will have been sufficient to drive down case and fatality numbers substantially, allowing for an almost complete lifting of restrictions. Although vaccination of children aged 12 years and over might have started by this point[112], other groups which have yet to be vaccinated might still suffer from relatively high incidence rates. As the oldest and most vulnerable population groups at highest risk of death from COVID-19 have been prioritised for vaccination, the overall fatality rate in the population and the health burden imposed by SARS-CoV-2 will decline. Hence, the perception of the remaining danger might be low: more than 10% of infected individuals are expected to suffer long-term sequelae of COVID-19 ("long-COVID") - symptoms of which can include shortness of breath, fatigue, and muscle weakness.[113-117]

Assuming increased international mobility due to, in particular, high vaccination coverage, a potential outbreak of a new VOC in one country may spread quickly to others. Without rapid intervention, increased mobility may result in simultaneous outbreaks across countries and regions - potentially putting healthcare systems under high pressure. In light of this danger, a joint effort of all European countries to prevent the emergence and circulation of VOCs seems crucial.[118, 119]

In short, countries with good access to vaccines and high vaccine uptake can, at worst, expect only modest waves of COVID-19 over the winter when maintaining moderate NPIs (e.g. no large indoor gatherings, face masks, physical distancing, good ventilation, and hygiene). In contrast, countries that have a lower level of vaccination coverage will experience more severe waves unless appropriate NPIs are implemented. Any new VOCs might challenge a successful mitigation or containment strategy, and in case of increased mobility, they are likely to spread quickly.

## The perspective for the coming 3-5 years

For the coming three to five years, the central questions are: Will we leave the pandemic behind? And if we do – when and how? To what degree will COVID-19 continue to play a role? Regarding the direct health impact of COVID-19, it is possible that it could become a disease that a child will encounter at a young age[120], acquiring a mild infection similar to contracting other coronaviruses. The time scale for this shift is uncertain. Early childhood exposure and recovery may help the immune system to protect the individual, should they encounter the virus again later in life, and should prevent them from experiencing severe symptoms. On the other hand, SARS-CoV-2 (and more so new VOCs) is more infectious and lethal than the known endemic human coronaviruses, and there is the continued risk of long-COVID. Similarities to Chikungunya suggest that long-COVID may become a great burden.[121] However, relief might come from new and improved post-exposure therapeutic options, such as antiviral medication and monoclonal antibodies.[122] Hence, there is mixed evidence whether SARS-CoV-2 will remain a serious threat to health in the long-term.

It is unclear whether eradication of SARS-CoV-2, i.e., a global reduction to zero incidence of infection[123], can be achieved. Global mass vaccination programs might only provide imperfect immunity to some individuals and will usually not reach certain subpopulations, leaving pockets of susceptibility. Transmissions within these subpopulations, the high proportion of asymptomatic COVID-19 infections, and waning of post-infection and vaccine-induced immunity could maintain the circulation of the virus in the global population. Even if eliminated in humans, the multitude of documented non-human hosts[124-127] suggest the virus could remain circulating with ongoing risks of infection of and potential further spread between susceptible human hosts. Furthermore, the virus could mutate within human or non-human hosts to escape immune response, potentially requiring repeated booster vaccinations. In any case, eradicating SARS-CoV-2 would require global political commitment and unified and uniform public assent that eradication is the overarching target. With the smallpox virus, the only virus able to infect humans to have been eradicated, a targeted and globally concerted approach over decades was necessary[128, 129], with a particular focus also on reaching deprived populations.[130]

Elimination, meaning here a temporary reduction to zero incidence of infection in one region or country through deliberate and continued measures, has been achieved in a small number of countries; e.g., Australia, China, New Zealand, Singapore, and Vietnam. With widespread vaccination, others may try to follow as elimination strategies can offer advantages over mitigation or suppression strategies with continued virus circulation.[131] Assuming that children will also be vaccinated, some of these countries might achieve high enough vaccine uptake to sustainably prevent local transmission. In other countries where immunity in the population is insufficient or too heterogeneous for elimination, SARS-CoV-2 is expected to remain prevalent at a comparatively low level, with recurring local and seasonal outbreaks.[120, 132] In the absence of eradication, epidemiological surveillance (and TTI) will need to remain in place and be further improved.[53] The level of immunity in the population will prevent widespread morbidity and mortality, but a significant danger might remain for unvaccinated vulnerable people.[4] A key societal question will be which level of such risk is deemed acceptable when balancing other societal goals.

Finally, Europe faces numerous indirect long-term impacts of the pandemic. Without intending to present a complete list, the consequences include:

**Health:** During the past year there has been a direct impact on healthcare services in regular care, particularly for patients with chronic conditions.[133-135] This includes reduced access to primary care[136], cancellation of elective medical and surgical procedures[137], and disruptions to screening programs.[138, 139] Potential suboptimal healthcare provision for non-communicable diseases might cause a progression of chronic diseases and complications of acute diseases. At-risk populations not sufficiently covered by screening programs might now develop serious disease within a 3- to 5-year period. Hence, further health- and economic burdens (increased sick days, decreased workforce, lost productivity, and increased healthcare costs) might be experienced by some countries due to the rise in the prevalence of non-communicable diseases.[140] With potentially increasing investment into pandemic preparedness, there is a risk of cuts in other public health sectors, aggravating the effects on prevention and chronic disease control. Additionally, the enormous consequences for mental health during this pandemic, especially in young people[75, 79], healthcare workers[141], and individuals already suffering from social disadvantage and discrimination[142-145], will have a protracted effect. Whilst the consequences do not appear to extend to higher suicide rates[146], there is the need to redirect services and ensure sound mental health and social care support to the population.

**Economy:** Although many facets of the economy in some wealthy countries may soon recover[71], others will struggle to overcome the economic crisis. The tourism industry has suffered gravely, endangering livelihoods and economies in countries that depend on it; and driving a widening divide between Northern and Southern Europe.[147] The cultural sector has also been hit economically by the pandemic.[148-151] Public debt has been growing, and this poses a risk to financial stability - especially in countries more strongly hit by the pandemic. Increasing digitalization, and remote and flexible work plans, will potentially change employment.[152] Meanwhile, the legislative and regulatory frameworks for these new forms of work, along with supporting mechanisms (e.g., for sound occupational health), are lagging behind.

**Society:** Inequalities have been exacerbated because of this pandemic.[144, 153, 154] This extends well beyond health inequalities[155] to gender[156, 157] and educational[158] inequalities. Many children have missed out on extended periods of face-to-face education, as well as general social interaction. At the same time, there has been further erosion of trust between citizens and states through a widening of the socioeconomic gap.[145, 159-161] These two factors present a threat to social cohesion and might cause social unrest in the years to come. Furthermore, the narrative of "outside threats" and "secure borders" in discussions about the virus might contribute to the intensification of pre-existing nationalistic and sometimes overtly xenophobic, social and political discourses.[162] The weakened cultural sector might further be challenged by long-lasting gathering restrictions, eliminating many platforms where communities could approach and engage with these issues. Moreover, a lot of progress on the Sustainable Development Goals, in particular on poverty reduction, will be reversed.[163]

Even if the rate of new infections eventually significantly decreases, the health-related, economic and social damages of the pandemic will be felt for a long time.

## The way forward

We can conclude that COVID-19 will continue to pose many challenges over the coming years. The economic, cultural, and health consequences of the pandemic are already immense and societies may need a long time to recover. The increasing availability of vaccines will bring significant relief over the next months, but if not accompanied with comprehensive strategies and public support they alone will not protect from further damaging outbreaks in the coming years. Limited uptake of vaccines and declining public adherence to NPIs impede the way out of the pandemic and in the worst case new VOCs can render current vaccines less effective.

The eradication, i.e. the complete global elimination, of SARS-CoV-2 seems unlikely. However, even if eradication cannot be achieved, strategies that aim to locally eliminate SARS-CoV-2 might be effective in some settings. If achieved, local elimination offers clear advantages over mitigation or suppression with continued virus circulation, at least until sufficient protection against severe symptoms is granted in the population. A successful strategy for elimination or suppression of SARS-CoV-2 would require a political commitment, unified and uniform public assent that elimination or the goal of low case numbers is the overarching target. To achieve said target, a clear, evidence-informed, and context-relevant strategy, as well as concerted efforts and actioning are crucial. Countries committing to that strategy would need to have (a) rapid vaccination programs across age groups, (b) sufficient NPIs that may only be lifted if the susceptible population at risk is small, (c) close communication between policymakers and a wide range of experts to weigh the societal costs and benefits of measures against each other, (d) mitigation of virus influx from regions with higher incidence, and (e) sufficient public health infrastructure. This infrastructure entails basic public health resources, well-trained personnel of sufficient number, well-functioning TTI systems, widespread sequencing of the virus variants, and well-established molecular surveillance mechanisms. International coordination and cooperation on all these points and on continued development of new drugs and vaccines (also for potential new VOCs) is essential.

In line with the Sustainable Development Goals, healthy lives should be a global common good and initiatives like COVID-19 Vaccines Global Access (COVAX) should receive more support. Support of low- and middle-income countries by high-income countries is not only crucial to mitigate VOCs, but is mandated by the principle of solidarity.[103, 164, 165] In the long-term, a global One Health approach to pandemic preparedness and control is crucial - respecting the interdependence of humans, animals, and the environment.[166]

# Discussion of parameters, strategies and their context

The following presents more detailed elaborations of some of the aspects discussed in the main text and a summary of important additional topics. For a more comprehensive narrative in each of these sections there is, inevitably, some overlap with previous text.

## Long-term strategy

To minimize the damage caused by the COVID-19 pandemic, a long-term strategy set on a common, global and overarching goal is required. By communicating a common goal that societies are working towards and by clearly formulating the reasoning behind the implementation of measures, they will be perceived as less arbitrary.[77] Such a strategy must be comprehensible and based on scientific evidence not only from epidemiology, but from a wide range of disciplines. Communication between politicians and experts for transparent, evidence-informed policymaking and comprehensive systematically updated context-relevant risk communication strategies is crucial. However, to be comprehensible a strategy also needs consistent concepts that are perceived as both understandable and fair. Hence, and vitally, any strategy needs to be underpinned by considerations of justice and (global) inequalities. The more comprehensible and fairer such models of pandemic management are, the more people will be willing to support more extensive interventions in their everyday life.[83] This also includes showing that not all population groups are affected by the pandemic in the same way.

Specifics of the strategy will necessarily vary locally and also change over time in the face of more data about (a) the virus, particularly current and newly-emerging VOCs, (b) the development of vaccines and treatments, and (c) the harms accrued to individuals, communities, and societies through restrictions. Any strategy needs to balance the damage of being harmed by the virus against the damage by the measures to contain it. This will shift in response to the vaccination progress. Thus, it would be problematic if governments became fixed upon a specific strategy and remained committed to it regardless of new evidence and circumstances.

Any strategy should not simply be developed by politicians and imposed on the public: such impactful strategies should, as far as possible, be based on societal consensus, although recognising that some politicians may base their views purely on ideological premises. Moreover, measures are much more likely to be successful if they are developed through a process of co-production with those who must implement them and who are most affected.[167]

## Vaccination coverage

**When will sufficient vaccination coverage be reached?**

Vaccination programs are progressing in Europe (see Figure 3).[38] The chances of achieving high vaccination coverage depend on political leadership, access to vaccines and concerns and anxieties in relation to vaccination.[168] The latter especially differs from country to country.[169] At present, with mostly the eldest and most vulnerable receiving the vaccine, vaccine uptake has been generally high.[41, 42] In the younger groups, willingness to be vaccinated is lower[43, 44], limiting the final average uptake. In some countries, only about 40% of the adult population currently plan to accept the offer of vaccination.[45] Moreover, it is concerning that, in some countries, there is significant vaccine hesitancy among healthcare workers.[46, 47] However, perception of increasing vaccine uptake might motivate those who are hesitant.[49]

If the aim is to reach population immunity, children will have to be vaccinated as well, because the required level of immunization for population immunity likely cannot be reached otherwise. If not immunized, infections in children might become central for an annual autumn or winter epidemic. High incidence in children also poses the risk that the virus may spread to vulnerable individuals in the general population with waning immunity. Children are likely to become eligible for vaccination in 2021.[112] However, parental perspectives on and ethical considerations around childhood vaccination may pose significant challenges.[170]

As of April 2021, vaccination programs in many countries have slowed down. Repeatedly changing policy recommendations and constant media coverage seem to have unsettled many people, after evidence of rare adverse, sometimes fatal, side-effects emerged mid-rollout for the AZD1222 (AstraZeneca) and Ad26.COV2.S (Johnson & Johnson) vaccines.[39, 40] Likely because of this, some people rather prefer to wait for a vaccine of their choice. At a later stage, increasing vaccination coverage and successful control of the pandemic may decrease the willingness to get vaccinated at all because the perceived risk of unwanted severe side-effects of vaccination might exceed the risk of contracting the disease.[171] This can be seen with other potentially-lethal infectious diseases. Once those who can and want to be vaccinated have been done, significant efforts may be required to encourage further people to become vaccinated. This would ideally be achieved through a coherent risk-communication strategy to effectively address the 'infodemic' and limit and address the circulation of inaccurate or misleading information about vaccines.

Despite these challenges, it is to be expected that most high-income countries will finish their first round of vaccination this year, whereas sufficient vaccination coverage in many low- and middle-income countries will take considerably longer. Widespread vaccine nationalism[172], underfunding[173] and patent laws[172] make the COVAX initiative function sub-optimally. With the current vaccines and manufacturing capacities, sufficient coverage for achieving population immunity in the poorest countries is not expected to happen before 2023. Thus, the production and global distribution of the vaccines must be increased massively and rapidly. Potential escape variants, arising from poorly controlled viral spread in countries without adequate vaccine access, or waning immunity might necessitate repeated vaccinations, further slowing down the process of global vaccination.

**Measures during vaccination rollout**

*Without careful containment and test-trace-isolate measures*, the population remains vulnerable to COVID-19 during the rollout of vaccination programmes. A lack of appropriate caution in the relaxation of restrictions will lead to high morbidity, with risk of long-COVID, and mortality. High incidence also favors the emergence of new variants, which can threaten the success of the vaccinations. However, there is an increasing pressure to ease measures as a larger fraction of the population has been vaccinated. As can be observed in the example of Chile, this can have grave consequences.[174] All public health policy responses to these demands should thus be well considered.

Immunity certificates or passports to enable the return to normal life for vaccinated, tested, or recovered people have been considered or introduced in some regions.[175, 176] These have significant ethical and social issues associated with them. The rules for any use of such immunity certificates (or similar) will have to be openly and thoroughly discussed regarding their immunological and ethical consequences, specifically in the light of escape variants and restricted availability of vaccinations.[177] The distinction between vaccinated and not (yet) being vaccinated could become another engine of inequality.

Furthermore, there is a need to reconsider the core metric for measuring the state of the epidemic: namely, incidence. Incidence denotes the number of positively-tested COVID-19 cases during a certain time interval normalized to the population. Many discussions or rules for implementing or lifting NPIs are guided by incidence thresholds. However, if more and more people become vaccinated, the infections will concentrate only in those groups of the population that are still susceptible, i.e., younger people. In this case, a low incidence would still mean a large number of cases in younger age groups.

For example, an incidence of 50 per million people per day could initially mean that 0·005% of under 30-year-olds were infected each day. If we then assume that a third of the population were under 30 years of age and the rest of the population was completely immunized, the same incidence would mean that 0·015%, thus three times more, of under 30-year-olds were infected each day. This incidence in the total population would then correspond to a three-fold higher incidence in those under 30 years old.

Keeping incidence thresholds for tightening and loosening measures as they are now will therefore put younger people more at risk, further burdening a group that has been severely affected by the pandemic, psychologically[75, 79, 178], economically[178], and educationally.[178, 179] On the other hand, younger people tend to be less risk-averse[180] and may be willing to take the risk in exchange for more individual freedom. Moreover, with increased vaccination among the elderly, the same incidence means a lower burden on hospitals and lower deaths. This means that current incidence thresholds would at a later stage correspond to lower risks to healthcare systems than they do now. A last aspect to consider on the matter of incidence is that the total incidence remains a rough measure of how well contact tracing can work, even after vaccination. As the feasibility of contact tracing should be a main factor for deciding incidence thresholds[106], this would be an argument against changing the thresholds. Nevertheless, this issue will need to be openly discussed with involvement of all stakeholders.

## Digital health systems and operations research to organize mass vaccinations

The delivery of vaccines and medical accessories involves complex supply chains, and the fragility of mRNA vaccines, which require a very good low-temperature cold chain, and may have to be stored at -20° to -80° Celsius, further complicates planning and logistics.[181] Countries with successful early vaccination programs during the COVID-19 pandemic, such as Israel and the United Kingdom, have benefited from an early start of mass vaccination and a steady vaccine supply:

Israel stands out for its national digital health network and electronic medical record system, which covers all citizens and can be accessed by all health management organizations (HMOs) in the country. The HMOs are independent and compete for members with a mix of public and private health care services, but a tight regulation and hierarchical structure in combination with the interconnected digital network allows the HMOs to implement a national health operation efficiently. Furthermore, organizational and logistic frameworks to facilitate the cooperation between government, hospitals and emergency care providers are well-established, and operations and health policy research, as well as digital health concepts are used to improve healthcare procedures.[181, 182] A detailed review of these and other factors which contributed to Israel's successful vaccination program has been provided by Rose and colleagues.[183]

Digital health systems also played a key role in the British vaccination program. As part of the prior operations research planning, optimal locations of vaccination centers were computationally estimated in a manner that ensured that every citizen could reach the nearest center within 10 miles from home.[184] For the supply chain management, a data analytics company was contracted to create a comprehensive supply database for vaccines, accessories and equipment.[185] The system also integrates information on trained staff for the vaccinations, non-identifiable patient data, and required materials in order to help prevent delays. Additionally, it provides up-to-date progress reports on vaccinations to the NHS to facilitate close monitoring. Further elements of the vaccination program that may have contributed to the early success in the UK have been discussed more comprehensively in a recent article by Baraniuk.[186]

Overall, many of the tools and strategies used in Israel and the UK in the areas of digital health management and analytics, as well as operations research, are transferable to other countries. Their deployment could help to increase the efficiency of vaccine delivery in settings with interdependent supply constraints.

## Engineering controls to reduce airborne transmission

There is unequivocal evidence that airborne spread is the dominant route of spread for SARS-CoV-2. Studies on human behaviors, practices and interactions in choir meetings, slaughterhouses, gyms and care homes have presented evidence consistent with airborne spread of SARS-CoV-2.[187] Long-range transmission between people in adjacent rooms but never in each other's presence has been documented in quarantine hotels.[188] Healthy building controls, such as better ventilation and enhanced filtration, are a fundamental - but often overlooked - part of risk reduction strategies that could have benefits beyond the current pandemic.[189]

Steps should be taken to ensure good ventilation in populated buildings to mitigate aerosol transmission. Priority should be given to spaces where ventilation is absent or inadequate, where there are several people in close proximity or for extended periods of time and those where infectious persons are more likely to be present. Optimizing natural ventilation by opening windows, increased air exchange in small rooms with low ceiling heights, scaling up the ventilation in high-occupant-density situations or in locations where masks are not worn all of the time are suggested.[109] Improving on this can become a global challenge since significant additional resources, not directly linked to healthcare budgets, will be needed. In addition, there has been limited guidance on specific ventilation and filtration targets. Notwithstanding, improved air quality in confined spaces may not only help to prevent infectious diseases well, but also to improve well-being and performance, e.g. learning in school children.

## Waning immunity

The duration of post-infection and vaccine-induced immunity to COVID-19 might show pronounced individual heterogeneity with some people not forming efficient immunity at all and others developing an immune response that might protect from reinfection for decades. Antibodies against SARS-CoV-2 have been shown at nine months post-infection.[22] About 95% of subjects retain immune memory at six months after infection.[23-25] However, reinfections have also been observed.[26-28] In some individuals reinfections are possible even just a few months apart.[190] Mechanisms for that as well as the expected average frequency of reinfection are not well known. In the case of SARS-CoV-1, humoral immunity was described to last for up to two to three years whereas antigen-specific T-cells were detected up to 17 years after infection.[191] It is important to keep in mind that circulating antibody levels are not necessarily predictive of T-cell memory or the level of protection. To conclude, waning immunity is a realistic risk and may necessitate booster shots in the years to come.

When they occur, reinfections are likely to be less severe because leftover baseline immunity may shorten the course of infection and dampen inflammatory responses. Antibody disease enhancement, analogous to what has been observed in Dengue fever[192], could in principle occur. However, no evidence so far exists that a reinfection will lead to more severe symptoms.

## Evolution of SARS-CoV-2

A key unknown in relation to the future of the pandemic is the ability of the virus to evolve in ways that increase its transmissibility, its disease severity, or its potential to escape from vaccine induced immunity. It was thought that the SARS-CoV-2 virus would evolve more slowly than other RNA viruses as it contains a proofreading mechanism. However, there has been a clear step change in emergence of constellations of mutations over time, termed "variants of concern". These often include specific mutations, for example, D614G, in the spike protein which enhanced binding to the ACE2 receptors on human cells.[193] This mutation is present in the currently important VOCs, including Alpha, Beta, Gamma and Delta. Another mutation, N501Y, involving a substitution of asparagine for tyrosine as the amino acid at position 501, allows the spike protein to bind more tightly to the ACE2 receptor, thereby further increasing the transmissibility of disease.[194] This mutation is also present in the VOCs Alpha, Beta and Gamma. A third mutation, E484K, reduces the ability of antibodies generated following vaccination or previous infection to bind to the spike protein.[58, 195] This mutation is present in Beta, Gamma and other variants under investigation.

The rollout of vaccination will inevitably change the environment within which the virus is circulating, creating an evolutionary pressure for further mutations against which existing vaccines may be less effective. However, many mutations do not increase the fitness of the virus and may even weaken it, for example by reducing the ability of the spike protein to bind to the receptor. Thus, much will depend on whether there is one or a small number of genotypes of the virus that are optimally configured for transmission. Research showing convergence of evolution of the spike protein in different SARS-CoV-2 lineages supports this possibility.[196]

This question has been addressed in an analysis of three variants of concern that have emerged in the pandemic, Alpha, Beta, and Gamma.[197] Martin and colleagues note that the same mutations have arisen independently in geographically dispersed populations, suggesting that, at least in some ways, the evolution of the virus may be converging on an optimally fit genotype.[197] However, they note that changes in the environment in which the viruses are being transmitted may create new opportunities. Variants bearing the N501Y mutation only began to emerge in the autumn of 2020. Having reviewed the evolution of the virus so far and of coronaviruses in other hosts, Martin and colleagues suggest that the most likely scenario is that the virus will evolve in ways that converge on one or more related "supervariants" with increased transmissibility and potential for vaccine evasion and they list a set of codons that such variants might be expected to possess.[197] However, it is not possible to exclude the situation in which other evolutionary pressures arise, particularly given the very short time during which this virus has been circulating in humans, and based on experience with other viruses.

## How to improve adherence to rules and recommendations

**Clearer communication**

As the assumed effectiveness of measures is a key predictor of their protective effects[84], it will remain critically important to improve scientific communication about them.[77] This is crucial because specific policies, such as the goal of very low incidence, require the understanding of complex underlying systems. Politicians and scientists must speak clearly and truthfully to the public, neither underplaying nor overplaying the risks associated with the pandemic or the effectiveness of interventions. Scientists with a public profile must be extremely mindful of demarcating personal opinion and interpretation from widely accepted scientific fact. Failure to do so risks undermining the very public health measures and campaigns (e.g. vaccination) that scientists are propagating.

The media also has a role to play. It is apparent that coverage has regularly been influenced by the ideological stance of the media outlet. In the United States, for example, conservative media outlets have been highly critical of those warning about the risks of COVID-19, such as Anthony Fauci, and have promoted conspiracy theories. Studies at an individual and area level have demonstrated associations between use of conservative media outlets, such as Fox News, and belief in conspiracy theories, reduced mask wearing, and lower reductions in mobility. Another study, using survey data from the United States and United Kingdom, found that intention to be vaccinated was associated with use of broadcast and print media (as well as support for Hilary Clinton in 2016 or the Labour Party in the United Kingdom) but not with social media, except in one study that asked about reliance on it for information, which found an association with reduced intention.[198]

**Empowering measures**

Adherence to public health measures can only be achieved if people have the capacity to do so.[199] This insight is supported by the fact that especially low adherence has been observed for people in precarious working conditions.[200, 201] We thus need to focus on making measures socially acceptable, and focus on mental health and ways to prevent, or at least relieve, social, economic, and psychological burdens associated with the pandemic. Helping people to cope with the situation and strengthening society will ultimately benefit adherence and ensure the effectiveness of measures.[202] Therefore, governments need to provide more support of multiple kinds (economic aid, more mental healthcare, social help etc.). It is of critical importance to support people from lower socio-economic backgrounds. Where possible, stress in parents and thereby in children should be reduced. This is especially since children have been hit particularly hard from a mental health perspective.[75, 79] It is also vital to make help accessible to those unfamiliar with the local language, those unable to apply for help (for instance, due to digital exclusion), and those unaware of support offers. Support must be directed at those residing in a country, not merely official citizens of it, to prevent the aggravation of existing inequalities.

**Physical, not social, distancing**

It should be stressed that restricting the virus does not necessarily mean restricting social interactions per se. Politicians and scientific advisors should pursue policies that actively animate community at a time of loneliness, depression, and anxiety; but in ways that remain in agreement with the important mission of driving down cases and fatalities of COVID-19. For instance, investment in urban public health is very important, from green spaces to small and safe community gatherings. For the latter, people should be encouraged to meet outside in small groups to have social interactions in a physically distanced way.[203]

## One Health

A One Health approach to disease control considers the interdependence of humans, animals and the environment with interdisciplinary thinking and measures.[204, 205] Such a holistic multi- and transdisciplinary approach is required because it is insufficient to only consider a human health perspective in our interconnected world. Animal reservoirs most likely play an important role in SARS-CoV-2 and other viral infections. This is certainly the case with respect to the origin of viral human pathogens, e.g. in bats, pigs or chicks.[206] One particularly relevant animal in this regard might be the bat as a source animal from which viruses can emerge that are resistant to high temperatures or fever in humans.[207] As SARS-CoV-2 is now a mainly human-to-human transmitted virus, it is not entirely clear how much other animals, such as household animals or farmed animals, play an important role. Several animals that have been in contact with infected humans have been tested positive for SARS-CoV-2; minks, dogs, domestic cats, lions and tigers.[124, 125, 127] We should monitor the appearance of SARS-CoV-2 in these and other species closely.

Due to animal reservoirs and because COVID-19 is most likely a zoonosis[208], human intrusion into the habitat of animals needs to be considered in the context of pandemics. Overexploitation and habitat destruction significantly increases the risk of newly emerging and rapidly spreading vectors and diseases.[209] Environmental factors that are relevant in this context include light pollution and deforestation, mainly driven by expansion of land for agriculture.[210] From a One Health perspective, it seems essential to reduce global land use for agriculture.

The connections between animal, human, and environmental health are complex and require systems thinking. More focus on this interconnectivity should be placed in education, to foster awareness of the importance of human actions on such large scales. As we move further into climate change, a range of serious health issues will become more common.[211-214] A One Health framework as part of a planetary and global health perspective to study and manage these will be helpful.[166, 215]

# Contributors

ENI, VP, SB, SBM were involved in conceptualisation, methodology, project administration, and writing. PK, MEK, CM, BP, ES were involved in conceptualisation. RB, PB, ACV, SC, TC, UD, EGI, EGr, CH, PH, PK, MEK, TK, JK, NL, HM, CM, MM, AN, MPe, EP, MPi, JR, ESc, AS, ESz, ST, SVG, PW were involved in providing content and writing - review & editing.

# Declaration of interests

PB was supported by the Epipose project from the European Union's SC1-PHE-CORONAVIRUS-2020 programme (grant agreement number 101003688), and consulting fees were paid to his institution by Pfizer and Pfizer Belgium. ACV was supported by the Ministry of Culture and Science of the German State of North Rhine-Westphalia and the German Federal Ministry of Education and Research. TC was supported by the European Union's Horizon 2020 research and innovation programme project PERISCOPE (grant agreement number 101016233). EGI was supported by the Luxembourg National Research Fund. EGr received fees from the German Board of Pharmacists for educational events on COVID-19 and is the president of the German Society for Epidemiology. MK was supported by ZonMw grants number 10430022010001 and number 91216062, and the European Union's Horizon 2020 research and innovation programme project CORESMA (grant agreement number 101003480). NL was supported by European Union's Horizon 2020 research and innovation programme project EpiPose (grant agreement number 101003688), and the Swiss National Science Foundation (project number 176233). MM is a member of UK Independent SAGE. MPi was supported by the UK Economic and Social Research Council (ESRC) [ES/S013873/1; ES/T014164/1], the UK Medical Research Council (MRC) [MR/S035818/1], and Wellcome Trust [209519/Z/17/Z; 106612/Z/14/Z]. BP is a member of the Austrian National Bioethics Commission, and the European Group on Ethics in Science and New Technologies, advising the Austrian Government and the EU Commission respectively. Other research projects in the lab of ESz are partly funded by Merck Healthcare KGaA. All other authors have no competing interests to declare.

# Acknowledgements

First and foremost, the facilitators would like to thank all the consulted experts and collaborators for their great contributions.
ENI, VP, SB, SBM were supported by the Max Planck Society. RB was supported by the University of Luxembourg. PB has received funding from the Epipose project of the European Union's SC1- PHE-CORONAVIRUS-2020 programme, project number 101003688. ACV has received funding from the Digital Society research program funded by the Ministry of Culture and Science of the German State of North Rhine-Westphalia. SC was supported by the University of Malta. TC has received funding from the European Union's Horizon 2020 research and innovation programme under grant agreement No 101016233 (PERISCOPE). UD was supported by the National Research Programme project VPP-COVID-2020/1-0008. EGI acknowledges funding support from the Luxembourg National Research Fund as part of the COVID-19 Fast-Track research project CovScreen (COVID-19/2020-1/14715687). MEK was supported by grants from The Netherlands Organisation for Health Research and Development (ZonMw), grant number 10430022010001, and grant number 91216062, and by the H2020 project 101003480 (CORESMA). TK was supported by the Wroclaw University of Science and Technology. JK has received funding from the European Research Council (ERC) under the European Union's Horizon 2020 research and innovation programme (grant agreement no. 724460). NL has received funding from the European Union Horizon 2020 research and innovation programme, project EpiPose (grant agreement number 101003688), and the Swiss National Science Foundation (project number 176233). HM was supported by the University of Minho. MPe was supported by the Slovenian Research Agency (Grant Nos. P1-0403 and J1-2457). MPi is currently supported by the UK Economic and Social Research Council (ESRC) [ES/S013873/1; ES/T014164/1], UK Medical Research Council (MRC) [MR/S035818/1], and Wellcome Trust [209519/Z/17/Z; 106612/Z/14/Z]. ESz acknowledges funding by the Polish National Science Centre OPUS grant no 2019/33/B/NZ2/00956 and SONATA-BIS grant no 2020/38/E/NZ2/00305. The remaining authors have no funding source to declare.

# Data availability

Not applicable.